\documentclass[referee]{aa}  
\usepackage{graphicx}
\usepackage{epsfig}

\usepackage{txfonts}
\begin{document}
 \title{Automated detection of gravitational arcs}
   \author{C. Alard
          \inst{1,2}}
   \offprints{C. Alard}
   \institute{Institut d'Astrophysique de Paris, 98bis boulevard Arago, 75014 Paris, France\\
              \email{alard@iap.fr}
         \and
             Observatoire de Paris, 77 avenue Denfert Rochereau, 75014 Paris, France.}
	    
    \abstract
      {} 
      {This paper presents a method to identify gravitational arcs or more generally
      elongated structures in a given image.}
      {The method is based on the computation of a local
      estimator of the elongation.
      The estimation of the local elongation proceed in two steps:
      first the local orientation of the structure is computed, then in the next
      step, a rotation is performed, and the marginal distributions
      are used to compute the elongation.
      This procedure allows the computation of the local elongation at each
      point in the image.
      Then, using a threshold on the elongation map the elongated
      structures are identified and re-constructed using connectivity criteria.}
 {Finally a catalog of elongated structures is produced, and the properties
      of each object are computed, allowing the selection of potential arc candidates.
      The final selection of the arc candidates is performed by visual inspection
      of multi-color images of a small number of objects.}
{This method is a general tool that may be applied not only to gravitational
arcs, but to all problems related to the mapping and measurement of elongated structures,
in an image, or a volume.}

	  
    \maketitle
    \section{Introduction}
     The new large scale surveys offers deep and well sampled views of large regions of the sky.
     These views contains such a large number of objects that automated method to find a specific
     type of objects is of great interest. In particular, the new release of the CFHTLS survey (see 
      {\it http://www.cfht.hawaii.edu/Science/CFHLS/} for a description of the project)
     presents, a large set of 1 square degree images, with a corresponding size of about 
     $20000 \times 20000$ pixels. Among a large number of interesting objects, this survey
     contains a number of gravitational arcs, typically one per image. Considering the huge
     number of pixels, finding theses arcs or arclets is a good task for an automated software.
     The identification of potential arcs in the image is mostly a search for elongated objects.
     Thus an automated method designed to find arcs should first produce a catalog of elongated
     objects and select possible arcs candidates among them. In the end the procedure should be 
     concluded by a visual inspection of a small number of candidates. This method is a new
     approach, former methods relied on filtering, for instance, Lerzer, Scherzer \& Schindler,
     (2004 \& 2005), used anisotropic filtering, while Starck, Donoho \& Candes (2003) used
     wavelet Filtering. Arc detections using catalogs from the Sextractor software (Bertin \& Arnouts
     1996), were also performed by Horesh {\it et al.} 2005.
    \section{The method}
    
    In general arcs or arclets appear in the image as very elongated structures, almost flat along
    the elongated direction and nearly as narrow as the PSF along the other. Other structures in the
    image may have similar elongations, but along the smaller dimension they are rarely as narrow as
    the PSF. Some spiral arms or a random coincidence of small and faint structure, may be confused
    with arcs, but most structures are not as narrow an arclet can be. Thus by selecting elongated and 
    narrow structures in the image, we should build a sample which contains a significant
    fraction of arcs. Once this sample is constructed the next step will be to characterize the arcs
    in the sample, with respects to the other contaminants, which can be real structures in the image,
    but also defects due to bright stars. The paper is organized as follows: in the first part,
    an estimator of the local elongation at small scale is introduced. The second part analyze the effect
    of the noise in the image on the estimation of the local elongation. And finally, some illustration
    of the practical implementation of the method will be given.
    
     \subsection{Reconstruction of the narrow elongated structures in the image}
     
      \subsubsection{Estimating the local elongation at small scale}
       At each point in the image defined by its coordinates $(x_0,y_0)$ we wish
        to estimate the local elongation at the scale of the PSF.
       The elongation is computed by analyzing the pixels in a window of size about a few
       times the effective size of the PSF around the
       point of interest. The first point in this analysis is to estimate the local
       orientation of the structure, by computing its second order moments. Then using
       the orientation defined by the moments, the local axis ratio is estimated using
       the 2 marginal distributions $I_Y$, and, $I_X$
       in the proper axis. To be more specific, the orientation
       is defined by the second orders moments:
       $$
          a = \int I(x_0+x_1,y_0+y_1) \ x_1^2 \ dx_1 dy_1
       $$
       $$
          b = \int I(x_0+x_1,y_0+y_1) \ y_1^2 \ dx_1 dy_1
       $$
       $$
          c = \int I(x_0+x_1,y_0+y_1) \ x_1  y_1 \ dx_1 dy_1     
       $$
       The coordinates $(x_1,y_1)$, are local coordinates with origin at the image coordinates $(x_0, y_0)$.
       The integration are performed in the interval $-M<x_1<M$, and $-M<y_1<M$.  
       The direction of the main axis is then defined by the angle:
       \begin{equation}\label{Equation2}
         \tan(\theta) = -{\frac {c}{-1/2\,b+1/2\,a-1/2\,\sqrt {{b}^{2}-2\,ab+{a}^{2}+4\,{c}^{2}}}}
       \end{equation}
       We perform the rotation with angle $\theta$ and center $(x_0,y_0)$, 
       in order to move from the original
       local coordinates $(x_1,y_1)$, to the proper coordinates $(x,y)$. In proper
       coordinates the direction of maximum elongation is along the $x$ axis. We then define
       The marginal distribution:
$$
 I_X(x)  = \int I(x_0+x,y_0+y) dy 
$$
$$
 I_y(y)  = \int I(x_0+x,y_0+y) dx
$$
       At each position in the image $(x_0,y_0)$, we define the following measurement
       of the local elongation:
       \begin{equation}  \label{eq:Q_def}
         Q(x_0,y_0) = \frac{1}{2 M} \frac{I_Y(0)}{\sup \ \left[I_X(x) \right]_{[-M < x < M]}}
       \end{equation}
       each direction, then locally, $I(x_0+x,y_0+y)=f(x) \ g(y)$ and we infer that:
       $$
         Q(x_0,y_0) = \frac{1}{2 M} \frac{\int f(x) \ dx}{\sup \ \left[f(x) \right]_{[-M < x < M]}} \ g(0)
       $$
       In this calculation it was assumed that the profile along the small axis
       $g(y)$ is normalized, thus that $\int g(y) \ dy$ =1.
       \subsubsection{Properties of the estimator}
       The first case of interest is an arclet with flat profile along the tangential direction.
       Then, $I_X$ is constant, and we have simply:
       $$
         Q(x,y) = g(0)
       $$
       Assuming that the profile $g(y)$ may be written as a generic function of the width
       of the profile $a$, $g(y)=\frac{1}{a} \ G \left (\frac{y}{a} \right)$
        with $G(u)$ symmetrical in u and thus maximum at $u=0$,
       we infer immediately that the estimator $Q(x,y)$ will be maximum at the center of the arc
       with a value $Q(x,y)=\frac{1}{a} \ G(0)$, which means clearly that $Q(x,y)$ will reach its maximum
       value at the center of the narrowest arcs. We may turn now to the general case, when the 
       distribution $I_X$ isn't flat. It is easy to find an upper bound for the integral on $f(x)$
       in $Q(x,y)$ by using the upper bound on $I_X$, since $\sup I_X= \sup \left[f(x) \right]$, we infer that:
       $$
         Q(x,y) \leq g(0)
       $$                       
       Thus this estimator will be always maximum on flat and narrow elongated structures.
       Any variation along the tangential direction may be considered as the consequence of
       a finite scale length along this direction. In particular if the scale length of the
       variation is smaller than $M$, the estimator $Q(x,y)$ will give directly an estimation
       of the axis ratio. Assuming:
       $$
        f(x) = B \ F \left(\frac{x}{b} \right)
       $$
       And $b<M$, then
       $$
        Q(x,y) = k \ b \ g(0) = \tilde k \frac{b}{a} G \left(0 \right)
       $$
       When $b$ is larger than $M$ the estimator $Q(x,y)$ is no longer the axis ratio.
       Actually the value of the main axis $b$ saturate at the value of the box size $M$.
       The limiting value $M$ for $b$ is reached when the distribution becomes flat or equivalently
       when the effective scale-length becomes infinite.
\subsubsection{Effect of the noise  on the estimator}
We may now evaluate the fluctuation of the estimator $Q(x,y)$ due to the noise in the image.
The noise on the estimation of the local axis ratio may be decomposed in 2 parts. First, given
an orientation of the local axis, there is some amount of noise due to the intrinsic noise
on the marginal distribution. This part will be computed first. But there is also another
source of noise (a priori un-correlated), which is due to the noise on the orientation $\theta$ 
of the local structure.
Assuming a variance $\sigma_Y$ for $I_Y$, and $\sigma_X$ for $\sup \left[I_X \right]$, the variance on $Q$ may
be expressed as:
$$
 \sigma_Q = \frac{I_Y^2}{I_X^2} \ \left( \frac{\sigma_Y}{I_Y^2} + \frac{\sigma_X}{I_X^2} \right)
$$
The variance $\sigma_Y$ of the marginal distribution $I_Y(y)$ scales like
$\frac{1}{{2 \ M+1}}$ with respect to the variance $N^2$ on the total flux in the box. The signal
will scale like the integral on the profile marginal distribution $g(y)$ in a bin of with
 1 pixel normalized by the total integral on $g(y)$. Since the arcs have about the width of
 PSF, and that the PSF may be modeled by a gaussian of width 4 pixels, we find that $I_Y(x)$ 
at the maximum of the arc will be about $1/4$
of the total signal $S$. Then:
$$
\frac{\sigma_Y}{I_Y^2} \simeq \frac{16}{2 \ M +1} \ \frac{N^2}{S^2}
$$
We turn now to the estimation of $\sigma_X$. Considering that the distribution probability
 of the noise has a partition function
$\Phi(u)$, and a probability distribution $\phi(u)$, the probability distribution
of the maximum of the distribution will be: 
$$
P_M(u)= (2 \ M+1) \ \phi(u) \ \Phi(u)^{2M}
$$
 Assuming a gaussian probability distribution
for the noise, we find that:
$$
P_M(u)= K \ exp(-u^2/\sigma^2) \ (1+erf(u/\sigma))
$$
This distribution
may be approximately modeled by a gaussian distribution with an offset $u_0$s: 
$P_X(u)=A \ \exp \left(-{\left (u-u_0 \right)^2}/{{\tilde \sigma}} \right)$. 
for $M \simeq 1.5 \ {\rm seeing} \simeq 7 \ {\rm pixels}$, we find that $\tilde \sigma \simeq 1/4 \ \sigma$. 
And finally we infer the signal to noise ratio the variance $\sigma_X$:
$$
 \frac{\sigma_X}{I_X^2} \simeq \frac{\left(2 \ M +1 \right) N^2}{4 \ S^2} 
$$
And eventually, taking $M=7$, for a well sample PSF,
the signal to noise ratio for the estimator $Q$ reads:
$$
 \frac{Q}{N_Q} = \frac{Q}{\sqrt{\sigma_Q}} \simeq   \ 0.45 \left( \frac{S}{N} \right)
$$
Which means that even for low signal to noise, $\frac{S}{N} \simeq 10$, the noise will
be only 20 \% of the signal. This is small compared to the usual dynamics on $Q$ itself,
which for our choice of parameter varies from about 1 to 3. 
      \subsubsection{Noise on the estimation of the local orientation}
       An important problem is to estimate the error due to the noise image that we make
       in the estimation of $\theta$. Most of the noise on the arclets is due to the fluctuation
       of the background, thus it will be assumed the amplitude of the noise is nearly constant.
       Lets call $\sigma$ the variance of the pixels due to the background noise, the relevant
       noise on the moments is then:
       $$
         <a^2> \simeq \sum_{i,j} <I_{i,j}^2> x^4 \simeq \sigma \int  x^4 \ dxdy = \sigma \ 2/5 \ M^5
       $$
        Similarly:
       $$ 
         <b^2> = \sigma \ 2/5 M^5
       $$ 
        And:
       $$ 
         <c^2> = \sigma \ 4/9 \ M^6
       $$ 
        In order to obtain the variance on $\theta$ it is necessary to 
       differentiate equation (\ref{Equation2}) with respect to ($a$,$b$,$c$), and to add the
        respective variances. The final result may be expressed using the two proper second order momentum
         ($\alpha$, and $\beta$) of the system. Since:
       $$
        a = \sin(\theta)^2 \ (\beta-\alpha) + \alpha
       $$
       $$
        b = \sin(\theta)^2 \ (\alpha-\beta) + \beta
       $$
       $$
        c = \frac{\sin(2 \theta)}{2} \ (\alpha-\beta)
       $$
 \\
       $$
        <\theta^2> \simeq \frac{\cos\left(2 \theta \right)^2}{\left(\alpha-\beta \right)^2} \frac{2 \ M^6}{5}
       $$
 \\
         A useful quantity is the average variance. 
       $$
        \bar{<\theta^2>} \simeq 1/5\, \frac {M^6 \sigma}{\left(\alpha-\beta\right)^2}
       $$ 
       Considering that the total noise in a window of size $M$ is:
       $$
        N^2 = \sigma \ (2 \ M +1)^2
       $$
       And that the proper second order momentum may be rewritten, as a function
       of the proper size $L_0$, $L_1$, and of the total flux in the window, $S$: 
       $$
        \alpha = L_0^2 \ S
       $$
       $$
        \beta = L_1^2 \ S
       $$
       It is possible to estimate the variance on $\theta$:
       $$
         \bar{<\theta^2>}  \simeq  \frac{1}{20} \, \frac {N^2 \ M^4}{S^2 \left(L_0^2-L_1^2 \right)^2}
       $$
       For a flat distribution like an arclet the size $L_0$ along the main axis
       is close to the size of the window $M$, while
       the other dimension is small and may be neglected. In this case the former expression
       reduces to:
       $$
         \bar{<\theta^2>}  \simeq  1/20\,  \frac {N^2}{S^2}
       $$
       Which means that the error on $\theta$ depends only on the signal to noise.
       For small signal to noise, typically $\frac{S}{N} \simeq 10$ the error on $\theta$ is
       only of about:
       $$
         N(\theta) = \sqrt{\bar{<\theta^2>}} \simeq \sqrt{1/200} \ \ {\rm radians} \ \ \simeq 1^o
       $$
       The error on $\theta$ has an effect on the estimation of the local elongation.
       The arclet will be represented by a  quadratic polynomial
       in proper coordinates:
       $$
          I(x,y) = F(a_0 \ x^2 + a_1 \ y^2)
       $$
       Assuming that the arclet is mis-aligned by an angle $d\theta$ with its proper
       axis, the intensity of the arclet now reads:
       $$
          I = F\left(\left[a_0 + (a_1-a_0) d\theta^2 \right] \ x^2 + \left[a_0 + (a_0-a_1) d\theta^2 \right] \ y^2  +2 \ (a_0-a_1) d\theta \ x y \right)
       $$
       In particular we are interested with the effect of this rotation on the scale of the distribution
       along its small axis. The distribution along a particular axis may be estimated by using the reduced
       one dimensional marginal distribution along this axis. Thus the effect of the rotation may be estimated
       by calculating the effect on the marginal distribution along the y axis:
       $$
        I_Y = \int I(x,y) dx
       $$  
       Which may be re-written by changing the integration variable:
       $$
         I_Y = \int F\left(\alpha u^2 + \beta y^2 \right) du 
       $$
       With $\beta$:
       $$
        \beta = a_1 \left(1+\left(1-a_1/a_0 \right) \ d\theta^2 \right)
       $$
       By introducing the scale length $L_0=1/\sqrt{a_0}$, and $L_1=1/\sqrt{a_1}$,
       and the new scale length of the marginal distribution, $\bar L_1 = 1/\sqrt{\beta}$,
       we obtain:
       $$
        \bar L_1 = L_1 \left (1+\frac{1}{2} \left(-1+\frac{L_0^2}{L_1^2} \right) \ d\theta^2 \right)
       $$
       As estimated in the former section, $d\theta^2 \simeq \frac{1}{200}$, and in in our choice
       of parameters, $\frac{L_0}{L_1} \simeq 3$ , we find that the relative difference between 
       $\bar L_1$ and $L_1$ is only about 2 \%. The effect of the rotation on the main axis smaller
       again, and thus the error on $\theta$ has a negligible effect on the estimation of 
       the axis ratio.
\subsection{Effect of local curvature}
 In this paper it was assumed that arclets or elongated structures may be de-composed
in small rectilinear bits at the scale $2 M$, which is basically a few times the size
of the PSF. This is an excellent approximation for larger arcs, and the estimator
$Q(x,y)$ is designed for this specific task. But it is interesting
to study the behavior of the estimator for smaller arcs. As usual we assume that
without curvature at local scale the profile can be de-composed along each direction:
$I(x_0+x,y_0+y)=f(x) \ g(y)$. Assuming that structure is curved locally and may be
approximated by a circle of radius $R$ and that the tangential direction is along the
x axis, the curvature introduce a small shift $\delta y$ of the profile along the $y$ direction.
To the lowest order this shift is:
$$
 \delta y = \frac{x^2}{2 R} 
$$
The shift $\delta y$ has a negligible effect on $I_X$, thus the effect on $Q(x,y)$ will
be related to the variation due to $\delta Y$ on $I_Y$ only. With curvature, $I_Y$
may be re-written:
$$
 I_Y(y) = \int f(x) \ g(y+\delta y) dx \simeq g(y) \int f(x) dx + \frac{d g}{d y} \int f(x) \ \frac{x^2}{2 R} dx
+  \frac{d^2 g}{d^2 y} \int f(x) \ \frac{x^4}{4 R^2} dx 
$$
 Assuming that the profile along the $y$ direction, $g(y)$ is symmetrical and that the derivative is continuous, we have
$\frac{d g}{d y} = 0$. For an arclet, the profile along the $x$ direction $f(x)$ may be approximated by a constant $f_0$ in the relevant interval. Then $I_Y(0)$ reads:
$$
 I_Y = 2 \ M \ g(0) \left(1 + \frac{M^4}{40 R^2} \ \left[\frac{1}{g} \frac{d^2 g}{d^2 y} \right]_{y=0} \right)
$$
We have now to evaluate the second derivative of the functional $g$ near
$y=0$. Expanding $g$ to order 2 in $y$, we have:
$$
 g(y) \simeq A \left( 1 + \frac{x^2}{a^2} \right)
$$ 
and subsequently:
$$
 I_Y \simeq 2 \ M \ g(0) \left(1 + \frac{1}{20} \left(\frac{M}{R} \right)^2  \left(\frac{M}{a}\right)^2  \right)
$$
Numerically, the perturbation introduced by curvature may be evaluated as follows: $M$ is about 1.5 times
the seeing, $S$, and since $a=2 \ \sqrt(\log(2)) S$, we have, $M \simeq 2.5 S$. And
finally we have:
$$
\Delta_C \simeq 1.95  \left(\frac{S}{R}\right)^2
$$
Basically, this means that the error is still acceptable (20 \%, about the noise level
for faint objects), when, R, is about 3 times the seeing.
   \section{Practical implementation.}
   \subsection{Filtering}
      Finding arcs is equivalent to look for structures that have nearly the size of the
      PSF along one dimension. There is no scale length smaller than the PSF in the image,
      except artifacts due to the noise. Thus any filter that damps the lower frequencies
      while damping somewhat the noise will improve the detectability of the arcs. Such filters
      are basically band-pass filter, centered at the PSF scale. An example of such filter
      is a Mexican-hat filter:
      \begin{equation} \label{Equation1}
       M(x,y)=\exp^{-\frac{x^2+y^2}{b^2} }-0.5 \exp^{-\frac{x^2+y^2}{2 b^2}}
      \end{equation}
      The value of the constant $b$ should estimated so that the filter
      width match the size of the PSF. It is clear that this 2-dimensional filter, will
      improve the detectability of structures at about the size of the PSF. But the real
      point of interest is the effect of the filter on a one-dimensional structure like an
      arclet. To estimate the effect of the filter on the arc, we
      may take advantage of the symmetry of the filter by making a local rotation around the
      filter center  such that
      the X-axis will coincide with the direction tangential to the arc.
       Assuming that
      the arc has a  a flat profile along the tangential direction, and a gaussian profile 
      along the narrow direction: $I(x)=exp(-x^2/a^2)$, we may calculate 
      the scalar-product with the mexican-hat
      defined in Eq (1). The amplitude of this scalar-product will be proportional to the
      amplitude of the arc after filtering. The scalar-product reads:
      $$
       P(\eta) = \pi \ \eta  \ \sigma^2 \left(\frac{1}{\sqrt{1+\eta^2}} - \frac{1}{\sqrt{2+\eta^2}}  \right)
      $$
      The functional $P(\eta)$ peaks at $\eta \simeq 0.838$, and decrease
      quickly for larger values of $\eta$. Thus it is clear that for maximum
      efficiency, given a scale $a$ for the PSF, the scale $\sigma$ of the filter
      must be: $\sigma \simeq 1.19 a$.
\subsection{Using the estimator $Q(x,y)$ to find elongated structures.}
\begin{figure}[]
\centerline{\epsfig{file=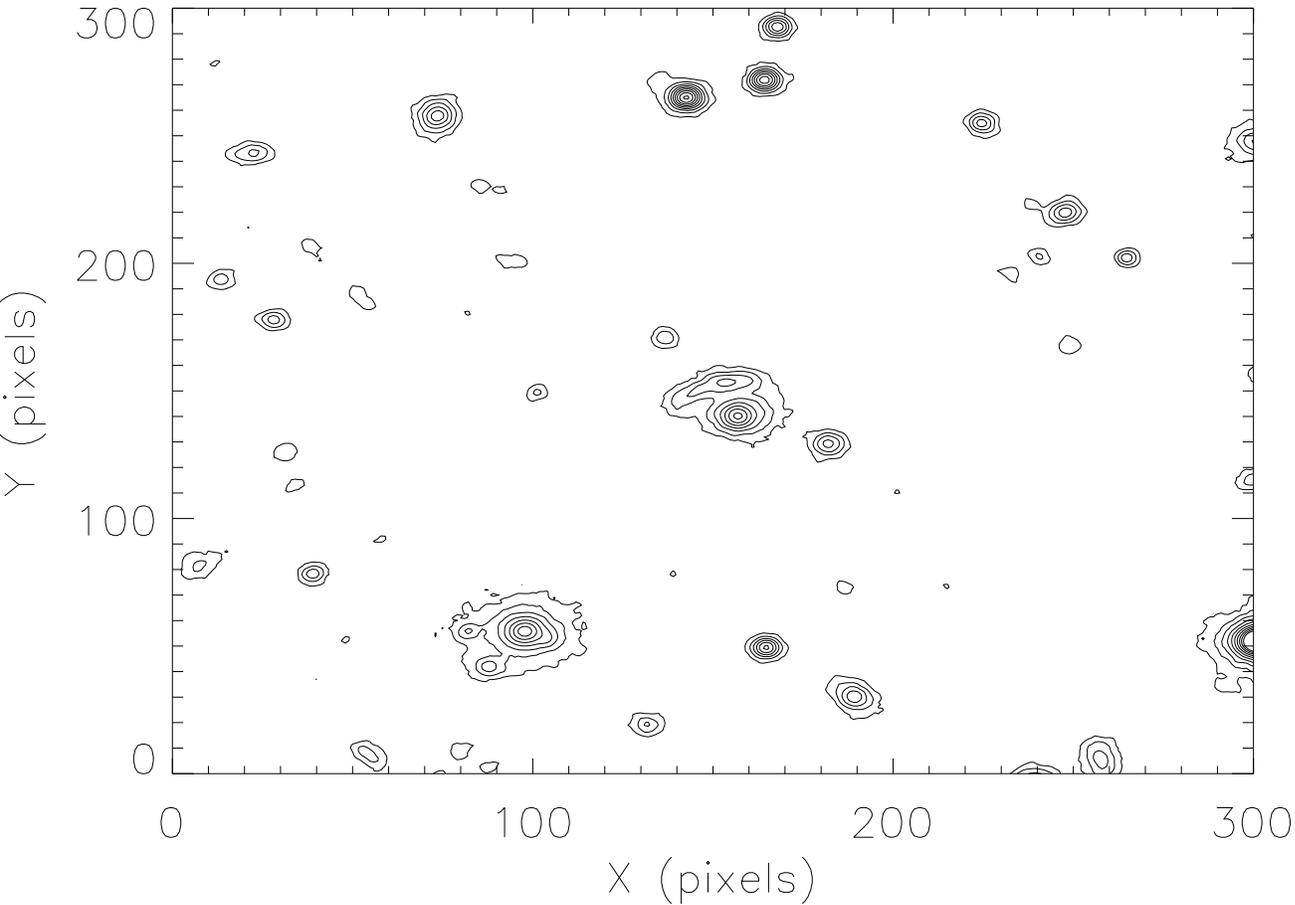,width=19cm}}
\caption{Contour plot of a CFHT image. 
In this figure an arclet is visible at the center of the frame,
near a bright source. This image will be used to illustrate the
computation of the local estimator $Q(x,y)$ - (see Fig. 2).}
\label{fig:c2}
\end{figure}

\begin{figure}[]
\centerline{\epsfig{file=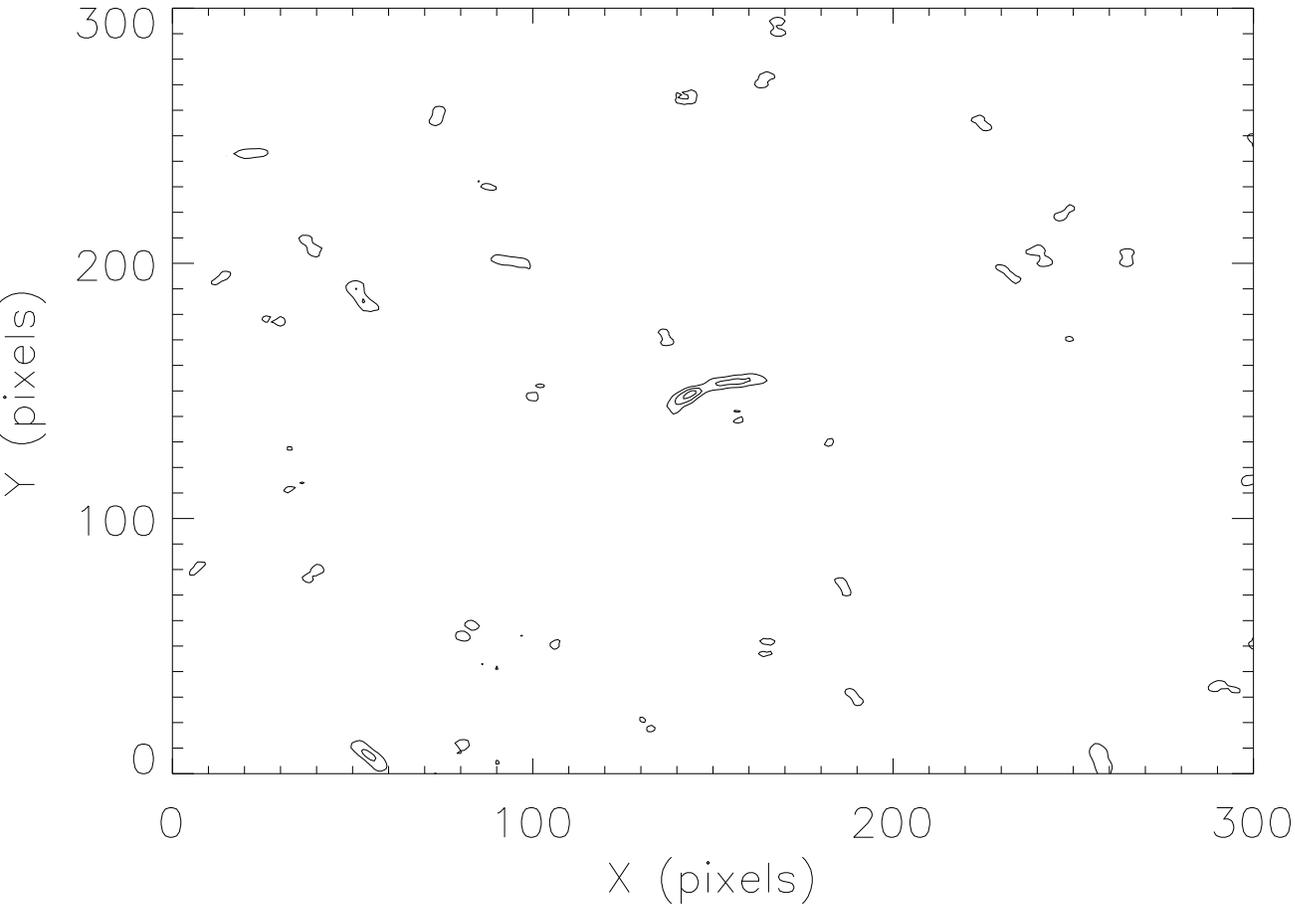,width=19cm}}
\caption{The contour plots in this image represents the local values of
the estimator $Q(x,y)$ at each point of the image presented in Fig. 1. The contours
were plotted using a linear scale, and the highest contours (which are located in the
arclet) are about 3 times higher than the lowest contours.}
\label{fig:c1}
\end{figure}

As explained in the former sections, at each point in the image the direction $\theta$ of
the local structure is computed using the second order moments, a rotation is performed in order
to be aligned with the direction of the local structure and finally the estimator $Q(x,y)$ is computed
using Eq. (\ref{eq:Q_def}). To illustrate the method, an image representing the value of the estimator
$Q(x,y)$ at each point of the image presented in Fig. (\ref{fig:c2}) is presented in Fig. (\ref{fig:c1}).
Local elongation of about 3 are reached within the arclet at the center of the image, and no-where else
in the image. Actually if larger images were taken, some objects other than arclets, may also presents
large value of the local elongation. For instance, edge-on spirals. Then the problem is to separate
the arclets from the rest of the sample. This can be done mostly by computing the properties of the local
structures as detected in Fig. (\ref{fig:c1}). The structures in the $Q(x,y)$ image may be reconstructed 
using the ink-blot technique. A structure is defined as a set of pixels larger than some value, and the
pixel must be connected each other. This way all the structures in the $Q(x,y)$ image may be
reconstructed,and their properties: number of pixels, length, curvature,...ect, may be measured. 
It is generally possible to separate most of the arclets form other structures using these 
parameters. For this particular task it may be also useful to measure some of the parameters of
the structures in the original image, like for instance their color or brightness. 
Table (1) gives the 5 largest structures found in the image presented in Fig. (\ref{fig:c1}).
Note that the actual field that was used is much larger than in Fig.  (\ref{fig:c1}), the
effective field was of 1000$\times$1000 pixels for Table 1.
\begin{table}
\begin{tabular}{|c|c|c|c|r|}
Number of pixels & mean value & pix nb/radius & radius & remarks \\
 \hline
 103 &  2.19  &     3.8  &   26.8 & Central arclet \\
 59 &       1.93 &       1.0 &      56.4 &  \\
 53 &       1.81 &      0.9 &       58.5 &  \\
 46 &       2.15 &      0.9 &       50.2 &  \\
 45 &       1.59 &      0.8 &       50.7 &  \\
\end{tabular}
\caption{paramaters of the structures found in the $Q(x,y)$ image (Fig. \ref{fig:c1}).}
\end{table}
\begin{table}
\begin{tabular}{|c|c|c|c|r|}
Number of pixels & mean value & pix nb/radius & radius & remarks \\
 \hline
 81   &  2.13   &  1.90   &  42.7   &         \\
 79   &  1.97   &  1.28   &  61.6   &         \\
 65   &  1.88   &  1.16   &  55.8   &         \\
 58   &  1.61   &  0.91   &  63.5   &         \\
 45   &  1.91   &  3.14   &  14.3   &     Small arclet   \\
\end{tabular}
\caption{paramaters of the structures found in the $Q(x,y)$ image.}
\end{table}
\begin{figure}[]
\centerline{\epsfig{file=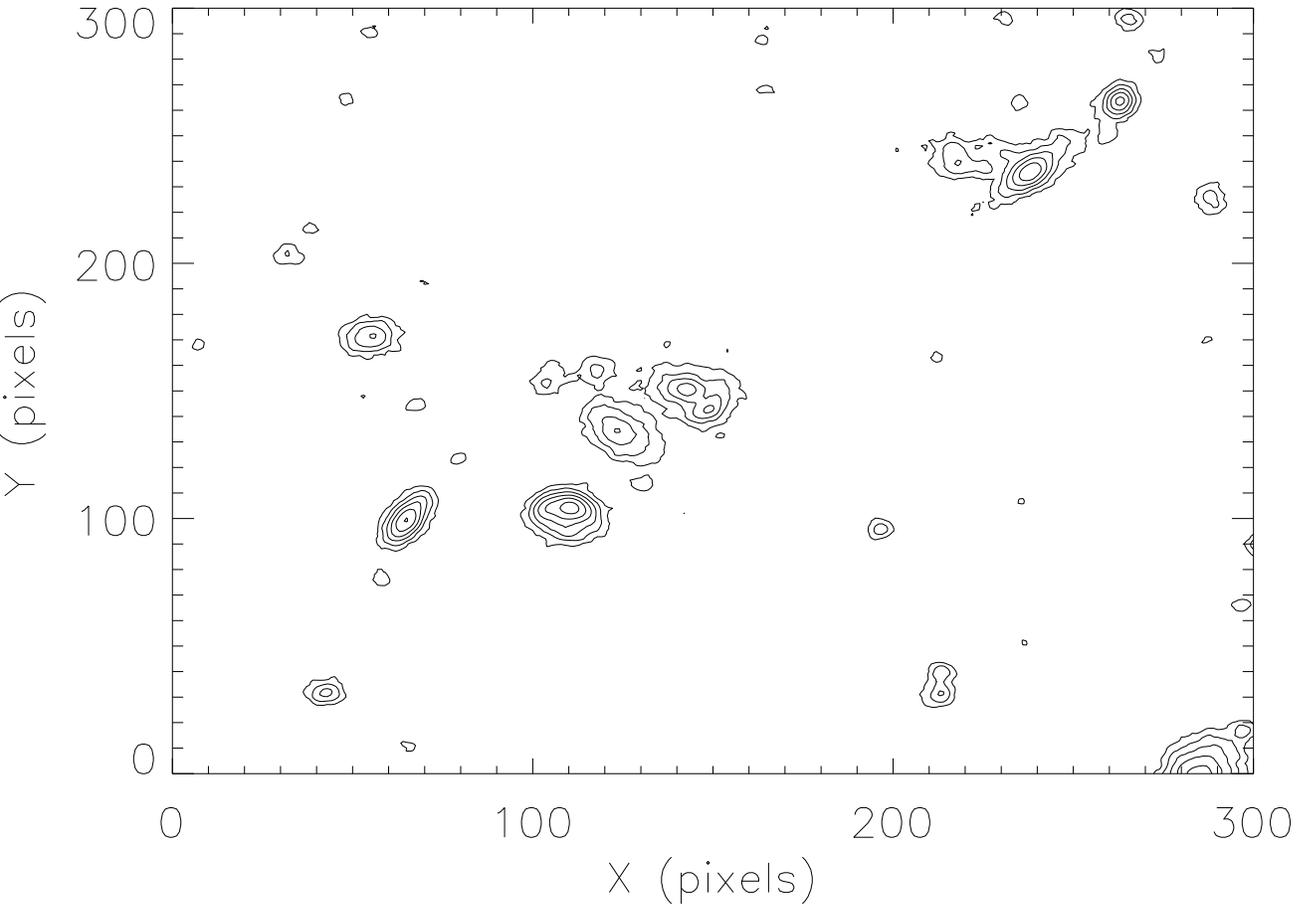,width=19cm}}
\caption{A small and faint arclet is located near the center of the image.
Note that this small arc is very close to the deflector, and that there is
some blending of the light with the lens.}
\label{fig:c4}
\end{figure}
\begin{figure}[]
\centerline{\epsfig{file=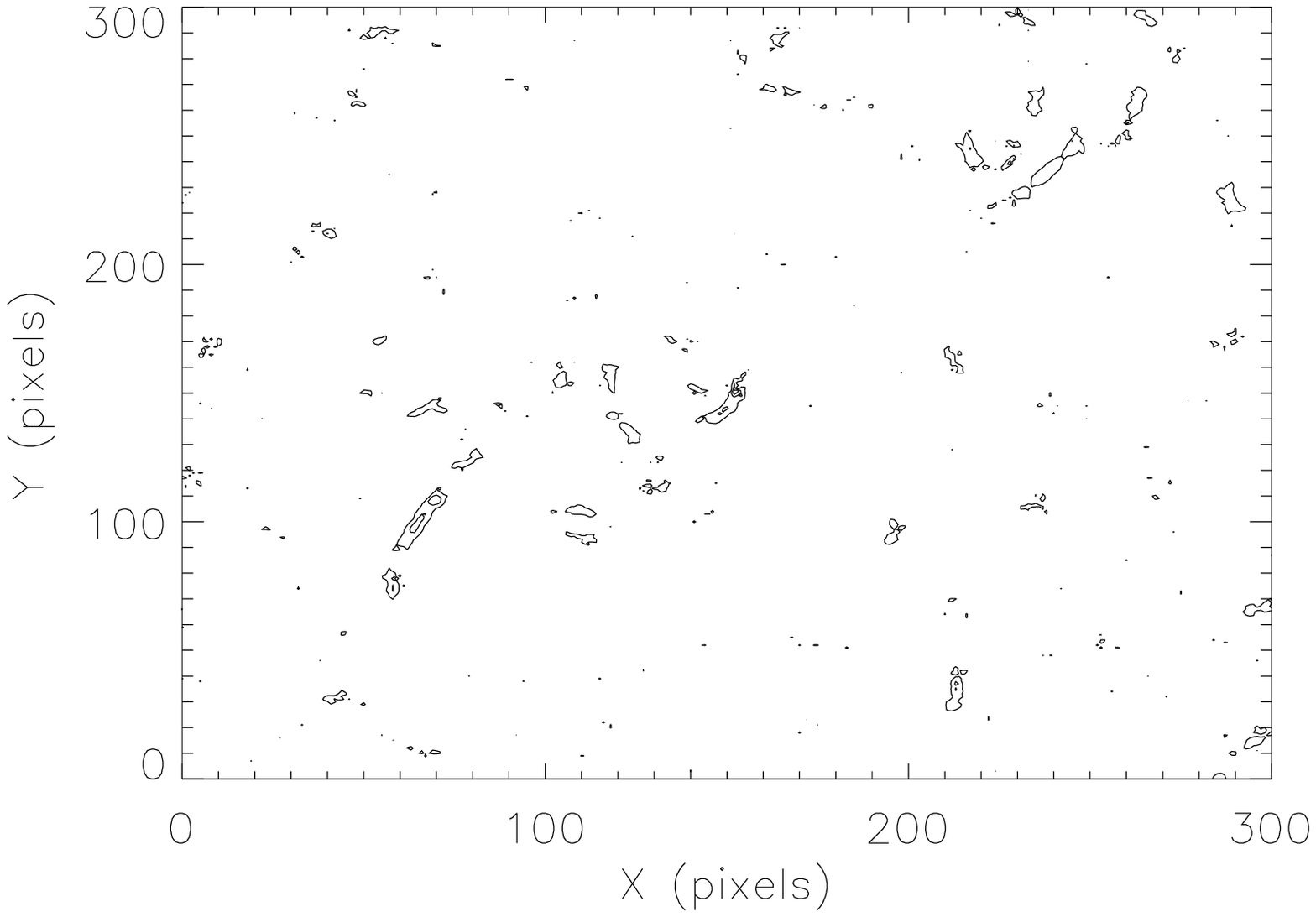,width=19cm}}
\caption{When estimating the local elongation $Q(x,y)$ the arclet appears now clearly,
and stands out in the image.}
\label{fig:c3}
\end{figure}

\subsection{Computing time}
Considering the size of the image, the computation of the estimator $Q(x,y)$ as to be fast.
The filtering and the computation of the second order moments can be greatly optimized
using well known techniques, like the decomposability of the filters, or partial buffering of
the calculations. However, it is more difficult to speed-up the calculation of the marginal distributions
 and of the estimator itself, since the orientation is not known in advance. 
But it is clear that the computation is not meaningful in all points of the image, first
it is possible to introduce a threshold in the filtered image to perform the computation. 
A further gain of computing time may be operated by calculating the estimator only near
uni-dimensional local maxima. Uni-dimensional local-maximas are maximas along a particular direction,
like for instance the kind of maxima that will be found on a direction perpendicular to an arclet.
Since a priori the orientation of the arclet is not known, the maximas will be searched both along
the $x$ and $y$ axis. Each local maxima along the $x$ or $y$ axis will be defined as a point where
the estimator $Q(x,y)$ has to be calculated. This procedure speeds-up the computation of $Q(x,y)$
by about a factor of 10 in a typical CFHTLS image.
\end{document}